\documentclass[conference]{IEEEtran}
\IEEEoverridecommandlockouts
\usepackage{cite}
\usepackage{amsmath,amssymb,amsfonts}
\usepackage{algorithmic}
\usepackage{graphicx}
\usepackage{textcomp}
\usepackage{xcolor}
\usepackage{color}
\usepackage{makecell}
\usepackage{subfigure}
\usepackage{comment}
\usepackage{url,graphicx,algorithm,algorithmic,multirow}

\newcommand{\cut}[1]{}
\def\BibTeX{{\rm B\kern-.05em{\sc i\kern-.025em b}\kern-.08em
    T\kern-.1667em\lower.7ex\hbox{E}\kern-.125emX}}

\begin{document}

\title{Spatial Co-location Pattern Mining - A new perspective using Graph Database}

\author{\IEEEauthorblockN{1\textsuperscript{st} Sanket Vaibhav Mehta}
\IEEEauthorblockA{\textit{School of Computer Science} \\
\textit{Carnegie Mellon University}\\
Pittsburgh, PA, USA \\
svmehta@cs.cmu.edu}
\and
\IEEEauthorblockN{1\textsuperscript{st} Shagun Sodhani}
\IEEEauthorblockA{\textit{Mila,} \\
\textit{Universit\'{e} de Montr\'{e}al}\\
Montreal, Canada \\
sshagunsodhani@gmail.com}
\and
\IEEEauthorblockN{2\textsuperscript{nd} Dhaval Patel}
\IEEEauthorblockA{\textit{Thomas J. Watson Research Center} \\
\textit{Yorktown Heights, NY, USA}\\
pateldha@us.ibm.com}
}
\maketitle



\begin{abstract}
Spatial co-location pattern mining refers to the task of discovering the group of objects or events that co-occur at many places. Extracting these patterns from spatial data is very difficult due to the complexity of spatial data types, spatial relationships, and spatial auto-correlation. These patterns have applications in domains including public safety, geo-marketing, crime prediction and ecology. Prior work focused on using the spatial join. While these approaches provide state-of-the-art results, they are very expensive to compute due to the multiway spatial join and scaling them to real-world datasets is an open problem. We address these limitations by formulating the co-location pattern discovery as a clique enumeration problem over a neighborhood graph (which is materialized using a distributed graph database). We propose three new traversal based algorithms, namely CliqueEnum\textsubscript{G}, CliqueEnum\textsubscript{K} and CliqueExtend. We provide the empirical evidence for the effectiveness of our proposed algorithms by evaluating them for a large real-life dataset. The three algorithms allow for a trade-off between time and memory requirements and support interactive data analysis without having to recompute all the intermediate results. These attributes make our algorithms applicable to a wide range of use cases for different data sizes.
\end{abstract}

\section{Introduction}
\label{sec:Introduction}
Google generates about 25 PB of data each day, significant portion of which is spatio-temporal data  \cite{vatsavai2012spatiotemporal}. NASA\cut{\footnote{http://data.nasa.gov/about/}} generates about 4 TB/day of spatial data. Now we have more spatial data than ever, both in terms of quantity and quality. Moreover, with the GPS enabled mobile and hand-held devices, we are able to capture richer geo-location data. Other sources include vehicles with navigation systems and wireless sensors \cite{yoo2014parallel}. These spatial datasets are considered nuggets of valuable information \cite{vatsavai2012spatiotemporal} and there is significant interest in extracting useful information for applications in geo-marketing, public safety and government services layout.

Spatial Data Mining \cite{tutorialSDM} is the process of discovering interesting and previously unknown, but potentially useful, spatial patterns from large spatial data. These spatial patterns include - spatial outliers, discontinuities, location prediction models, spatial clusters and spatial co-location patterns. \cut{Extracting these useful patterns from spatial data is more difficult than extracting such patterns from traditional numeric and categorical data due to the complexity of spatial data types, spatial relationships, and spatial auto-correlation.}
\cut{
\begin{figure}
\center
\includegraphics[width=1\linewidth]{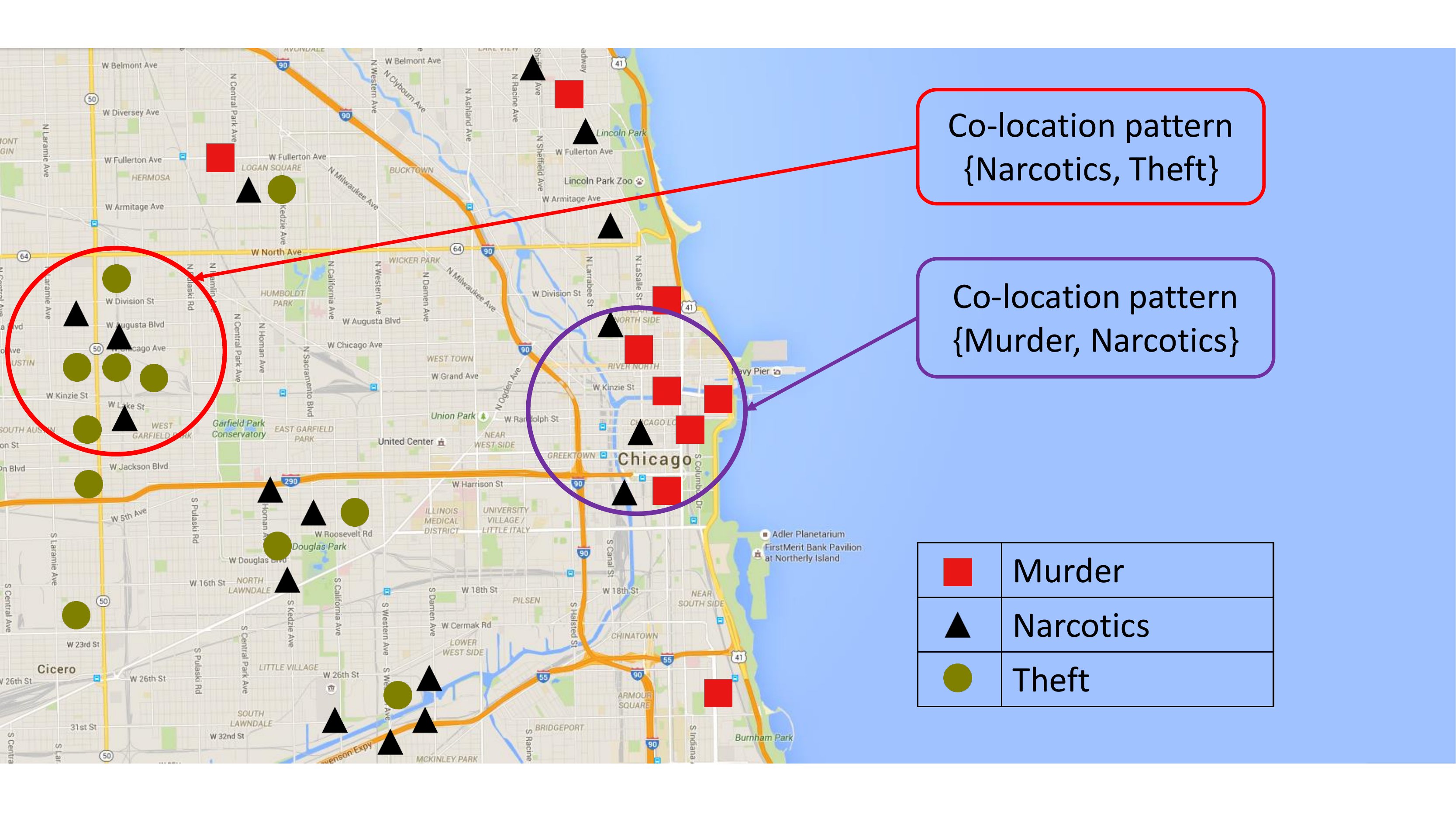}
\caption{SCPs illustrations over a real spatial crime dataset.}
\label{fig:ColocationPattern}
\end{figure}}
Our work in this paper focuses on mining one such pattern i.e., spatial co-location pattern, defined as \cut{co-location pattern\footnote{Spatial co-location pattern and co-location pattern are used interchangeably in this paper.}} a set of features that co-occur at many places \cite{shekhar2001discovering} . For example, in public safety\footnote{For better understanding of concepts related to co-location patterns, we use crime data for explanations and discussions.}, the co-location pattern $\{Murder, Narcotics, Theft\}$ indicates that these three crimes co-occur at many places. \textbf{Spatial Co-location Pattern (SCP) Mining:} Given a set of spatial features and their instances, spatial neighborhood relation and a prevalence threshold, spatial co-location pattern mining finds a set of prevalent co-location patterns. \cut{Figure \ref{fig:ColocationPattern} shows co-location patterns illustration of a real crime dataset consisting of three features - Murder, Narcotics and Theft.} \cut{\color{red}Murder crime instances are represented by square markers, Narcotics crime instances are represented by triangle markers and Theft crime instances are represented by circle markers.}

\cut{\color{red}\textbf{Applications:} Co-location pattern mining finds applications in various domains like location-sensitive advertisements \cite{yoo2007spatial}, weather prediction, public safety, geo-marketing, transportation, tourism and crime prediction \cite{talkColocation}. For example, consider a logistics company that provides services like registration,  training, cataloging, packaging etc. The requests for those services may be coming from different locations. Now business analysts for the company may be interested in discovering what types of services are requested by geographically nearby customers, in order to target these customers with location-sensitive advertisements and sell other services that the company offers.}  

\begin{figure}
\center
\subfigure[Basic outline of existing approaches]{

\includegraphics[width=0.9\linewidth]{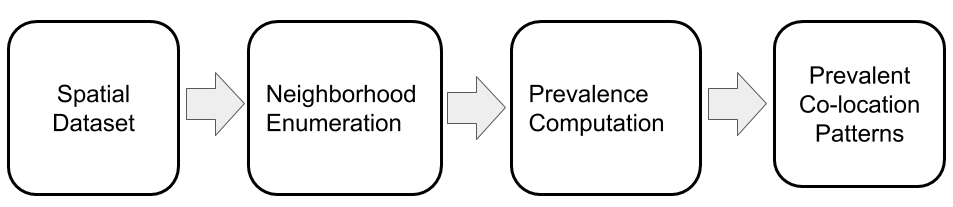}\label{fig:ExistingApproaches}}

\subfigure[Basic outline of our proposed approach]{
\includegraphics[width=0.9\linewidth]{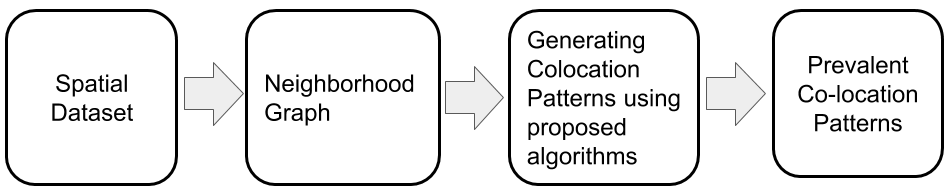}\label{fig:ProposedApproach}}
\caption{Basic outline of various approaches}
\end{figure}

Majority of approaches for SCP mining \cite{shekhar2001discovering,xiao2008density,yoo2004partial,yoo2005join,zhang2004fast,huang2003mining} have the common approach described in Figure \ref{fig:ExistingApproaches}. \cut{In Figure \ref{fig:ExistingApproaches},}Step 1(4) denotes the input(output). Step 2 (neighborhood enumeration) deals with exploration of neighbors in the spatial domain by multiway joins. Since current approaches use relational databases, they suffer from join pain. Step 3 deals with prevalence computation, where \textit{prevalence} is a metric used to ascertain the interestingness of the discovered patterns \cite{shekhar2001discovering} and it is a computationally expensive step. Steps 2, 3 are iterative in nature and results generated from previous iteration are used in next iteration. The enormous amount of data demands efficient techniques for computation, storage and retrieval of intermediate results. Most approaches, except \cite{yoo2014parallel}, are not distributed in nature and scaling out is a major challenge. Moreover, any modification in the neighborhood relation requires a complete re-computation rendering the previous computations useless. As a solution to these issues, we propose leveraging distributed storage and parallel data processing techniques for efficiently mining co-location patterns. 


With the advent of distributed graph databases, there is a scope of exploring SCP mining with graph databases by leveraging graph properties to perform efficient neighborhood enumeration. We propose an idea to bring the problem of SCP mining to the graph domain by modeling spatial data as a property graph. We show (in Section \ref{sec:Preliminaries}) that SCP mining is equivalent to clique enumeration on the property graph. We term this property graph as the \textit{neighborhood graph}. By choosing a distributed graph database to realize the neighborhood graph, we develop several techniques for SCP mining. Further, graph based models enables dynamic neighborhood relations as well. In this work we show application of graph databases to discover SCPs, thus, establishing a new and promising field for further explorations to other spatial data mining patterns.

\section{Related Work}
\label{sec:RelatedWork}
Approaches for discovering SCPs in the literature can be categorized into two classes, namely apriori algorithm based approaches and non-apriori algorithm based approaches. 

Apriori algorithm based approaches \cite{koperski1995discovery, shekhar2001discovering, yoo2004partial, yoo2007spatial, yoo2005join, munro2003complex, huang2003mining, huang2004discovering} focus on creating transactions over space on the basis of spatial relationships (e.g., proximity, etc). \cut{\cite{koperski1995discovery} introduces the problem of discovering co-location rules from spatial databases.}Shekhar et al.  \cite{shekhar2001discovering} proposes a \textbf{join-based approach} for SCP mining which uses a hybrid method of geometric and combinatorial approaches to perform neighborhood enumeration. They use generalized-apriori to identify candidate SCPs and prune them using the prevalence threshold based on the apriori property. The major bottleneck of this algorithm is the join step which makes it computationally expensive. After this, different algorithms \cite{yoo2004partial, yoo2005join, yoo2007spatial, xiao2008density, zhang2004fast} were proposed to improve upon the efficiency.

Yoo et al. \cite{yoo2004partial} proposes the \textbf{partial join-based approach} to overcome the limitation of the join step. \cut{They transactionize continuous spatial data while keeping track of the instances not modeled by transactions. Transaction-based apriori is used for transactions and adopts the join-based approach for residual instances not identified in transactions, thus, reducing the number of joins required.} Still the worst-case complexity of their approach is equivalent to the join-based approach. Yoo et al. \cite{yoo2005join} proposes the \textbf{join-less approach}, which eliminates the necessity of joins by using an instance look-up scheme. \cut{They materialize the neighborhood relationships through star neighborhood partition model in which star instances for each candidate SCP are generated. These instances form the super-set of SCP instances and final ones are generated from these star instances without any need of joins.} While the introduction of star instances avoids joins, but the generation of final SCP instances from them remains a major bottleneck of their approach as in worst case scenario all the star instances of all the sizes need to be checked for probable SCP instances. \cut{\color{red} These instances form the super-set of co-location pattern instances. Then co-location pattern instances are generated from star instances by using star neighborhoods without any need of joins. The approach also uses coarse level filtering on star instances to prune candidate co-location patterns without actually generating co-location pattern instances. The bottleneck step in this approach is the generation of co-location pattern instances from the star instances. In the worst case, the filter at the star instance level fails and all the star instances of all the sizes need to be checked for probable co-location pattern instances and all these star instances turn out to be co-location pattern instances}

In all of the above works, \textit{neighborhood constraint} is defined by a distance threshold which is the maximal distance allowed between two instances or events for them to be considered as neighbors. Qian et al. \cite{qian2009mining} proposes a greedy algorithm for SCP mining with dynamic neighborhood constraint.

Arunasalam et al. \cite{arunasalam2005striking} classifies spatial relationships into four different types - Positive, Negative, Self-Co-location, and Complex. To discover SCPs based upon complex relationships\cut{\color{red} (first introduced by Munro et al. \cite{munro2003complex})}, Verhein et al. \cite{verhein2007fast} proposes non-apriori algorithm based approach. Mohan et al. \cite{mohan2011neighborhood} defines a new type of co-location pattern termed as regional co-location pattern and  proposes a neighborhood graph based approach to mine such patterns.

We also use the concept of Neighborhood Graph but unlike \cite{mohan2011neighborhood}, we are interested in enumerating cliques of different sizes over this neighborhood graph. Bron et al. \cite{bron1973algorithm} is a well-known algorithm for finding all maximal cliques of an undirected graph. \cut{Some of its variants, including Dasari et al. \cite{dasari2014maximal} propose solutions for Maximal Clique Enumeration using Hadoop and MapReduce.}While we cannot directly use Bron et al. (or its variants as we do not need to generate all maximal cliques), we can use heuristics like how to partition the graph.

\cut{
\begin{table}
\caption{Description of the proposed algorithms
\label{table:proposedalgo}}
\begin{center}
\begin{tabular}{|c|c|c|}
\hline 
\thead{Algorithm} & \thead{Enumerating Candidate \\ Clique Instance}  & \thead{Validating Candidate \\ Clique Instance} \\
\hline
$CliqueEnum_G$ & Graph Traversal & Graph Traversal \\
\hline
$CliqueEnum_K$ & Graph Traversal & Key-Value Store \\
\hline
$CliqueExtend$ & Key-Value Store & Graph Traversal \\
\hline
\end{tabular}
\end{center}
\end{table}}
\textbf{Contributions:} In this paper, we build on the work of \cite{shekhar2001discovering} and consider the positive type of spatial relationship, clique type of co-location pattern. Our contributions are three-fold :
\begin{enumerate}
\item We model SCPs in graph domain by leveraging the concept of \textit{Neighborhood Graph} based upon the property graph model. We model SCP as a clique in neighborhood graph and formulate co-location pattern discovery as clique enumeration problem in neighborhood graph.
\item We present a vertex-centric algorithm to efficiently construct a neighborhood graph, given a spatial dataset and the neighborhood relationship (in the form of threshold distance). Our modeling and construction of neighborhood graph supports dynamic neighborhood relationship (more details in Section \ref{sec:Results}). 
\item We present three new algorithms\cut{(briefly described in Table \ref{table:proposedalgo})} \textbf{CliqueEnum\textsubscript{G}}, \textbf{CliqueEnum\textsubscript{K}} and \textbf{CliqueExtend}. The proposed algorithms are based on neighborhood graph traversal, are iterative in nature and follow apriori property. 
\end{enumerate}

\section{Preliminaries}
\label{sec:Preliminaries}
In this section, we recall concepts from SCP mining literature \cite{shekhar2001discovering} and introduce the neighborhood graph.

\textbf{Basic Concepts:} Let $F$ be a set of $k$ boolean spatial features $F = \{f_1, f_2, \cdots, f_k\}$. In case of crime database, as shown in Table \ref{table:dataset}, we have $F = \{Murder, Narcotics, Theft\}$. Let $D$ be a set of $n$ feature  instances, $D = \{d_1, d_2, \cdots, d_n\}$, where each feature instance is given by a 3-tuple $<identifier, feature, <latitude, longitude> >$. In Table \ref{table:dataset}, we have 10 feature instances and each feature instance is a 3-tuple. For example, $d_1$ is $<M.1, Murder, <lat1, lng1> >$. 

Two feature instances $d_i$ and $d_j$ are neighbors in the spatial domain if they satisfy the \textit{neighborhood relation}. We define the neighborhood relation in terms of the great-circle distance or orthodromic distance.\cut{\footnote{Orthodromic distance is defined as
\\
$Dist(d_i, d_j)^2 = 2r \arcsin(sqrt\{\sin^2\left(\frac{d_j.lat - d_i.lat}{2}\right) + \\ \cos(d_i.lat) \cos(d_j.lat) \sin^2\left(\frac{d_j.lng - d_i.lng}{2}\right)\}) $
where r is the radius of earth, $lat$ is latitude and $lng$ is longitude.}} So two feature instances $d_i$ and $d_j$ satisfy neighborhood relation if $Dist(d_i, d_j) \leq R_\delta$ and $d_i.feature \not= d_j.feature$. $R_\delta$ is defined as the distance threshold and is a domain specific constant. For a feature instance $l$, we define a neighborhood set, $N$, as a set of feature instances $I =\{i_1, i_2,\cdots, i_n\}$,  $\forall i_k \in I$, $l$ and $i_k$ are neighbors. 

A SCP is a subset of spatial feature set $F$. We have $I = \{i_1, i_2, \cdots, i_n\}$ as the \textit{row instance} of a SCP, $C = \{f_1, f_2, \cdots, f_n\}$, of size $n$ if $i_j$ is an instance of feature $f_j(\forall j \in 1, 2, \cdots, n)$ and $\forall x, y \in I, x$ and $y$ are neighbors. For a SCP, $C$, \textit{table instance} is the collection of all its row instances. The \textit{participation ratio, pr(C, $f_i$)}, for a feature $f_i$ of a SCP, $C$, is defined as the fraction of instances of $f_i$ that participate in any row instance of $C$. Formally, \begin{equation}\label{eqn:participationRation}
pr(C, f_i) = \frac{|distinct(\pi_{f_i}(all \; row \; instance \; of \; C))|}{|instances \; of \; feature f_i|}
\end{equation} where $\pi$ is a relational database projection operation. The \textit{participation index} of a SCP, $C$, is defined as $min_i\{pr(C, f_i)\}$. 

\textbf{Neighborhood Graph:}
We model the neighborhood relation, associated with feature set F, as a property graph $G=<V,E>$ where each vertex in \textit{V} is a  feature instance from \textit{D} and each edge in \textit{E} is a pair of vertices from $V$ satisfying the neighborhood relation. We term this property graph as the \textit{Neighborhood Graph}. An edge between vertices corresponding to feature $f_i$ and $f_j$ is labelled as $f_i:f_j$ if $f_i.feature \prec f_j.feature$.

We define $CandidateClique_C = \{v_{i_1}, v_{i_2}, \cdots, v_{i_n}\}$ as \textit{candidate clique instance} of a SCP, $C$, in neighborhood graph if $\forall i \in \{1, 2, \cdots, n\} $, $v_{i}.feature == f_{i}$ and $\exists$ an edge between every consecutive pair of vertices in $CandidateClique_C$ and also between $v_{i_1}$ and $v_{i_n}$. 

We define $Clique_C = \{v_{i_1}, v_{i_2}, \cdots, v_{i_n}\}$ as \textit{clique instance} of a SCP, $C$, in $G$ if $Clique_C$ is $CandidateClique_C$ and $v_{x}$ and $v_{y}$ are connected by an edge, $\forall v_{x}, v_{y} \in  Clique_C$. We state the following lemma without proof.
\\
\textbf{Lemma 1:} Clique instance in neighborhood graph is equivalent to a row instance of a SCP.

\begin{table}
\centering
\caption{Sample crime dataset
\label{table:dataset}}
\begin{tabular}{|c|c|c|}
\hline 
Feature Instance ID & Feature & Location \\
\hline
M.1 & Murder & $<lat1, lng1>$ \\
\hline
N.1 & Narcotics & $<lat2, lng2>$ \\
\hline
T.1 & Theft & $<lat3, lng3>$ \\
\hline
W.1 & Weapon Violation & $<lat4, lng4>$ \\
\hline
M.2 & Murder & $<lat5, lng5>$ \\
\hline
N.2 & Narcotics & $<lat6,lng6>$ \\
\hline
T.2 & Theft & $<lat7, lng7>$ \\
\hline
W.2 & Weapon Violation & $<lat8,lng8>$ \\
\hline
M.3 & Murder & $<lat9, lng9>$ \\
\hline
M.4 & Murder & $<lat10, lng10>$ \\
\hline
\end{tabular}
\end{table}

\begin{algorithm}
\begin{algorithmic}[1]
\caption{Neighborhood Graph Construction}
\label{alg1}
\REQUIRE Spatial Dataset $D$, Distance Threshold $R_\delta$
\ENSURE Neighborhood Graph $G(V, E)$
\STATE $V = \{\}; E = \{\}$
\FORALL {feature instance $d_i \epsilon D$} 
\STATE $v' = createVertex(d_i)$
\STATE $V = V \cup \{v'\}$
\ENDFOR 
\FOR{vertex $v \epsilon V$} 
\STATE $V' = searchNeighbors(v)$ 
\FOR{vertex $v' \epsilon V'$} 
\STATE $e = createEdge(v, v')$
\STATE $E = E \cup \{e\}$
\ENDFOR
\ENDFOR
\end{algorithmic}
\end{algorithm}
\begin{figure}
\center
\includegraphics[width=1.0\linewidth]{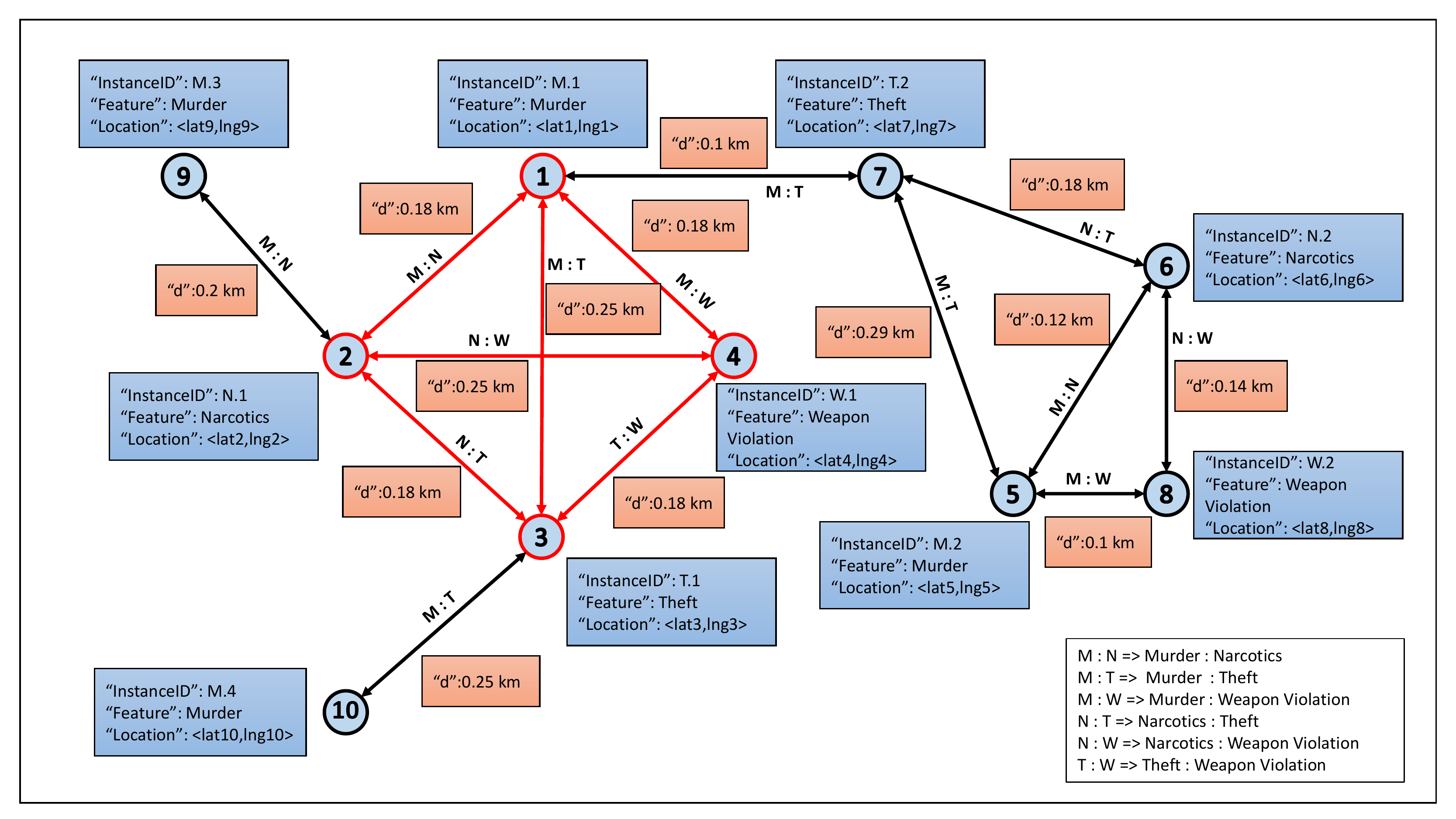}
\caption{Neighborhood graph corresponding to dataset as shown in Table \ref{table:dataset}}
\label{fig:neighborhoodgraph}
\end{figure}
\section{Neighborhood Graph Construction}
\label{sec:NeighborhoodGraph}
Let $G(V,E)$ be a neighborhood graph which we materialize using a graph database. As, $G$ is an instance of the property graph model, we have property-value pairs assigned to both vertices and edges in $G$. Algorithm \ref{alg1} shows the steps involved in constructing the $G$. The input is a spatial dataset $D$ and neighborhood relation (in terms of distance threshold $R_\delta$). Figure \ref{fig:neighborhoodgraph} is the neighborhood graph constructed for spatial crime dataset shown in Table \ref{table:dataset}. There are three main steps:

\textbf{Step 1:} Vertex creation and insertion (Line 2-5): This step initializes all the vertices for $G$. It uses the $createVertex$ method which takes a feature instance $d_i$ from $D$ as the input and returns a vertex $v'$ such that value of each attribute corresponding to instance $d_i$ in D becomes a property corresponding to the vertex $v'$ in $V$. For example, we have as instance $d_1$ as $< M.1, Murder, <lat1, lng1> >$ of type $<identifier, feature, location>$ and corresponding $v'$ has 3 property-value pairs "InstanceID": M.1, "Feature": Murder and "Location": $<lat1,lng1>$ as shown in Figure \ref{fig:neighborhoodgraph}.

\textbf{Step 2:} Neighborhood Exploration (Line 7): This step finds all the pair of vertices that satisfy the neighborhood relation\cut{$Neighbors: (V, V) \rightarrow Boolean$ over the space $S$}. We have $Neighbors(v_i , v_j ) \rightarrow True$ if both the vertices satisfy the Euclidean Norm i.e., $\|v_i.location - v_j.location\| \leq R_\delta$ where ``location" is a property of vertices. It uses the $searchNeighbors$ method which takes a vertex $v$ as input and returns a set of vertices $V'$ such that each vertex
$v'$ in $V'$ is a neighbor of $v$.

\textbf{Step 3} Edge creation and insertion (Line 8-11): This step generates all the edges for the graph. It uses the $createEdge$ method which takes vertices $v$ and $v'$ as input and either returns an edge instance or returns null. It returns an edge if $v.feature \prec v'.feature$ where "feature" is a property of vertices. In this case, the edge is labeled as $v_i.feature: v_j.feature$. Also the distance between $v$ and $v'$ is set as a property of the corresponding edge. Otherwise null is returned. A partial ordering is defined among the features. For our sample database we use lexicographic ordering.

\section{Methodology}
\label{sec:Methodology}
We discussed about neighborhood graph construction, $G$, and also saw that a row instance of a SCP is equivalent to a clique instance in $G$. So enumerating all row instances of a SCP is equivalent to enumerating all clique instances on $G$. Enumerating clique instances works in two steps:
\begin{enumerate}
\item Candidate clique instance enumeration.
\item Candidate clique instance validation.
\end{enumerate}
We propose three algorithms to generate prevalent SCPs. These algorithms differ in terms of how the above two steps of enumerating clique instances gets executed.
\begin{enumerate}
\item \textbf{CliqueEnum\textsubscript{G}} - Enumerate candidate clique instances for size-$k$ SCPs based upon the traversal on $G$ and then validate these candidates for clique instances using traversal on $G$. 
\item \textbf{CliqueEnum\textsubscript{K}} - Enumerate candidate clique instances for size-$k$ SCPs based upon the traversal on $G$ and then validate these candidates for clique instances using size $k-1$ clique instances.
\item \textbf{CliqueExtend} - Enumerate candidate clique instances for size-$k$ SCPs by extending size $k-1$ clique instances and then validate these candidates for clique instances using traversal on $G$. 
\end{enumerate}
\begin{algorithm}
\begin{algorithmic}[1]
\caption{CliqueEnum\textsubscript{G} Algorithm}
\label{alg2}
\REQUIRE Neighborhood Graph $G(V, E)$, $minPrev$, $k$
\ENSURE $prevalentColocations$ - a key-value store where the value for key $k'$ corresponds to an ordered set of prevalent co-locations of size $k'$
\STATE $prevlanetColocations[1]$ = new $OrderedSet()$
\FOR{vertex $v \in V$} 
\STATE $prevalentColocations[1].insert(v.feature)$
\ENDFOR 
\FOR {$k'$ in $(2, 3, \cdots, k)$}
\STATE $candidateColocations =$
\item[] $aprioriGen(prevalentColocations[k'-1])$
\STATE $prevalentColocations[k']$ = new $OrderedSet()$
\FOR{$candidate \in candidateColocations$}
\STATE $candidateCycles = getCycles(G, candidate)$
\FOR{$cycle \in candidateCycles$}
\IF{$isClique(G, cycle)$}
\STATE $store(cycle)$
\ENDIF
\ENDFOR
\IF{$prevalance(candidate) \geq minPrev$}
    \STATE $prevalentColocations[k'].append(candidate)$
\ENDIF
\ENDFOR
\ENDFOR
\end{algorithmic}
\end{algorithm}
\subsection{CliqueEnum\textsubscript{G} Algorithm}
\label{subsection:cliqueEnumG}
CliqueEnum\textsubscript{G} is a fully traversal based algorithm as both candidate clique enumeration and validation steps are traversal on $G$. Explanation of the detailed steps of Algorithm \ref{alg2}:

\cut{\textbf{Line 1} Instantiate the ordered set for key $k'$ = 1. }\textbf{Line 2-4} Set of size $1$ (singleton) co-locations is just the set of features, $F$, and the SCP instances are the vertices in the $G$. The participation index of all singleton SCP is $1$ so all of them are prevalent by default. \cut{\textbf{Line 5-20} Loop through $k'$ = $2,\cdots,K$ to generate prevalent SCPs for each size.}

\textbf{Line 6} Set of candidate SCPs is generated using the \textit{aprioriGen} method (as described in \cite{agrawal1993mining}). Failure to generate prevalent SCPs leads to early termination of the algorithm. As mentioned earlier, a partial ordering is maintained when labeling edges in $E$. The same ordering is used when generating candidate SCPs (to avoid redundant computations).
\cut{For example, consider Figure \ref{fig:neighborhoodgraph}. Feature set for this neighborhood graph is $F = \{Murder,Narcotics, Theft, Weapon Violation \}$. Partial ordering, based on lexicographic order, can be defined as $ \{ Murder \prec Narcotics \prec Theft \prec Weapon \\ Violation\} $. Suppose prevalent co-location patterns of size-2 are $\{Murder, Narcotics\}$, $\{Narcotics, Theft\}$ and $\{Murder, \\Theft\}$. So candidate co-location patterns of size-3 are $\{Murder, \\Narcotics, Theft\}$. $\{Murder, Theft, Narcotics\}$ and the remaining 4 permutations will not be generated as size 3 candidate prevalent co-location patterns due to partial ordering constraint.}

\cut{\textbf{Line 7} Instantiate the ordered set for key $k'$.}
\textbf{Line 8-9} For each candidate SCP, a set of candidate clique instances is generated. These candidate clique instances correspond to cycles in the $G$ and are enumerated by traversal over the $G$. For a SCP, $C = \{f_{i_1}, \cdots, f_{i_k}\}$, a graph traversal query of the form: $G.getVertices(``feature", f_{i_1}).traverseEdges(f_{i_1}$:$f_{i_2})\cdots$ $.traverseEdges(f_{i_1}:f_{i_k}).filter(based \; on \; starting \; vertex)\\.path(instanceid)$ is executed. So start traversing $G$ from vertices with feature $f_{i_1}$, then move along the edges labeled as $f_{i_1}:f_{i_2}$ to reach vertices with feature $f_{i_2}$ and continue traversing until we encounter vertices with feature $f_{i_k}$. Then traverse along the edges labeled as $f_{i_1} : f_{i_k}$ to reach vertices with feature $f_{i_1}$ and then filter the path traversed till now on the basis of the starting vertices so that we enumerate all size-$k$ cycles for the given SCP. We leverage $path()$ operator to keep track of the traversal. \cut{so in the above case path (instanceid) keeps track of the path traversed in the form of instanceid of the vertices participating in the path.} \cut{For example in Figure \ref{fig:neighborhoodgraph}, sample graph traversal query for candidate co-location pattern $\{Murder, Narcotics, Theft\}$ is as follows:

$G.getVertices("feature", Murder).traverseEdges(Murder: Narcotics).traverseEdges(Narcotics:Theft)\\
.traverseEdges(Murder:Theft).filter(based\ on starting \\vertex).path(instanceid)$

So candidate clique instances for candidate co-location pattern \\$\{Murder, Narcotics, Theft\}$ are: $\{M.1, N.1, T.1\}$ and \\$\{M.2, N.2, T.2\}$.}
Further, the edges were labeled when constructing $G$ and the underlying graph database indexes the labels, making this traversal query very fast. This query is triggered by the $getCycles$ method.

\textbf{Line 10-14} For each SCP instance (represented as a cycle in $G$), we validate if the cycle forms a clique by traversing over $G$ again\cut{and checking if an edge exists between all pair of vertices. } (this traversal is executed by $isClique$ method). A short circuit condition is placed for size-2 and size-3 as all edges and triangles are trivially cliques. 
\cut{In above example, we obtained $2$ candidate clique instances for size-$3$ co-location pattern $\{Murder, Narcotics, Theft\}$ and as size-$3$ candidate cliques are trivially cliques so we have $2$ clique instances for this co-location pattern at the end of this step.} In \textbf{line 12}, we store cliques in form of set of unique instances of each feature type occurring in the co-location. \cut{These counts are later used in line 15 to calculate the prevalence for the co-location.}

\textbf{Line 15-17} computes prevalence of candidate SCP using the set of unique instances of each feature type (which were saved in line 12). \cut{For candidate co-location pattern C = $\{Murder, \\ Narcotics, Theft\}$, we compute participation ratio for each feature - $pr(C, Murder) = \frac{2}{4} = 0.5, \; pr(C, Narcotics) = \frac{2}{2} = 1.0$ and $pr(C, Theft) = \frac{2}{2} = 1.0$.
\\
Participation index of $C = min\{0.5, 1.0, 1.0\} = 0.5$. So if we have $minPrev \geq 0.5$ then C is prevalent co-location pattern of size-3.}
Clique Enumeration (Line 8) and Clique Validation (Line 11) can be executed in parallel and Algorithm \ref{alg2} can be scaled horizontally, provided the underlying storage supports execution of queries in parallel.
\begin{algorithm}
\begin{algorithmic}[1]
\caption{CliqueEnum\textsubscript{K} Algorithm}
\label{alg3}
\REQUIRE Neighborhood Graph $G(V, E)$, $minPrev$, $k$
\ENSURE $prevalentColocations$ - a key-value store where the value for key $k'$ corresponds to an ordered set of prevalent co-locations of size $k'$
\STATE $prevlanetColocations[1] = new OrderedSet()$
\FOR{vertex $v \in V$} 
\STATE $prevalentColocations[1].insert(v.feature)$
\ENDFOR
\STATE $Mcurrent$ = new $KeyValueStore()$ 
\STATE $Mprevious$ = new $KeyValueStore()$ 
\FOR {$k`$ in $(2, 3, \cdots, k)$} 
\STATE $candidateColocations =$
\item[] $aprioriGen(prevalentColocations[k'-1])$
\STATE $prevalentColocations[k']$ = new $OrderedSet()$
\STATE $Mcurrent.clear()$
\FOR{$candidate \in candidateColocations$}
\STATE $candidateCycles = getCycles(G, candidate)$
\STATE $Mtemp$ = new $KeyValueStore()$ 
\FOR{$cycle \in candidateCycles$}
\IF{$validateClique(Mprevious,cycle)$}
    \STATE $Mtemp.insert(cycle)$
\ENDIF
\ENDFOR
\IF{$prevalance(candidate) \geq minPrev$}
    \STATE $prevalentColocations[k'].append(candidate)$
    \STATE $add(Mtemp, Mcurrent$)
\ENDIF
\ENDFOR 
\STATE $Mprevious = Mcurrent$
\ENDFOR
\end{algorithmic}
\end{algorithm}
\subsection{CliqueEnum\textsubscript{K} Algorithm}
\label{subsection:cliqueEnumK}
CliqueEnum\textsubscript{K} is a partial traversal based algorithm as only the first step - candidate clique enumeration involves traversal on $G$. Second step, validating candidate cliques, is performed by looking up a key-value store which stores clique instances for size $k-1$ SCPs. For size $k-1$ clique instance, the key is defined to be the first $k-2$ vertices and the value is the last vertex. \cut{Since this defines a one-to-many mapping (multiple clique instances will have the same key, but last vertex id), we store a list the last vertex ids corresponding to each key.} Explanation of the detailed steps of the Algorithm \ref{alg3}:

\textbf{Line 2-4} These are same as lines 2-4 for Algorithm \ref{alg2}. \textbf{Line 5-6} Two key-value stores instantiated to store clique instances. $Mcurrent$ stores the clique instances validated in the current iteration while $Mprevious$ stores the clique instances validated in the previous iteration and used for validating cliques in the subsequent iteration. 

\textbf{Line 7-9} These are same as lines 5-7 for Algorithm \ref{alg2}. \cut{\textbf{Line 10} $Mcurrent$ is cleared before storing clique instances for the current iteration.} \textbf{Line 11-25} These are similar to lines 8-19 for Algorithm \ref{alg2} with two major modifications. First, the candidate clique instances of size $k$ are validated using the clique instances of size $k-1$ stored in $Mprevious$ (line 15). Consider Table \ref{table:DataStructure} for the following example. We have candidate SCP $X = \{Murder, Narcotics, Theft, Weapon\; Violation\}$ under consideration. For this SCP, we get candidate clique instances as $\{M.1, N.1, T.1, W.1\}$ using traversal on $G$ (as mentioned in line 12). For validating this candidate clique instance we look for $<Key: \{M.1, N.1\}, Value: T.1>$ and $<Key: \{M.1, N.1\}, Value: W.1>$ clique instances in the key-value store corresponding to SCP $\{Murder, Narcotics, Theft\}$ and $\{Murder, Narcotics, Weapon \;Violation\}$ respectively. As both key-value pairs are present, candidate clique instances $\{M.1, N.1, T.1, W.1\}$ forms a clique. This logic is encoded in the $validateClique$ method. A short circuit condition is used for size-2 and size-3 clique instances just like in Algorithm \ref{alg2}.

The second modification being the clique instances are stored in $Mtemp$ in the form of key-value pairs as demonstrated in Table \ref{table:DataStructure}. \cut{These instances are added to $Mcurrent$ in line 21, if the SCP, to which the instances correspond to, is prevalent. The store $Mcurrent$ itself is copied over to $Mprevious$ in line 24.} 
Further, clique enumeration (line 12) and clique validation (line 15) can be executed in parallel provided the underlying storage supports execution of queries in parallel and Algorithm \ref{alg3} can also be scaled horizontally. 

\begin{table}
\centering
\caption{Table showcasing efficient way of storage of clique instances corresponding to the Neighborhood Graph shown in Figure \ref{fig:neighborhoodgraph}
\label{table:DataStructure}}
\includegraphics[trim= 0cm 2cm 0cm 0cm,width=1.0\linewidth]{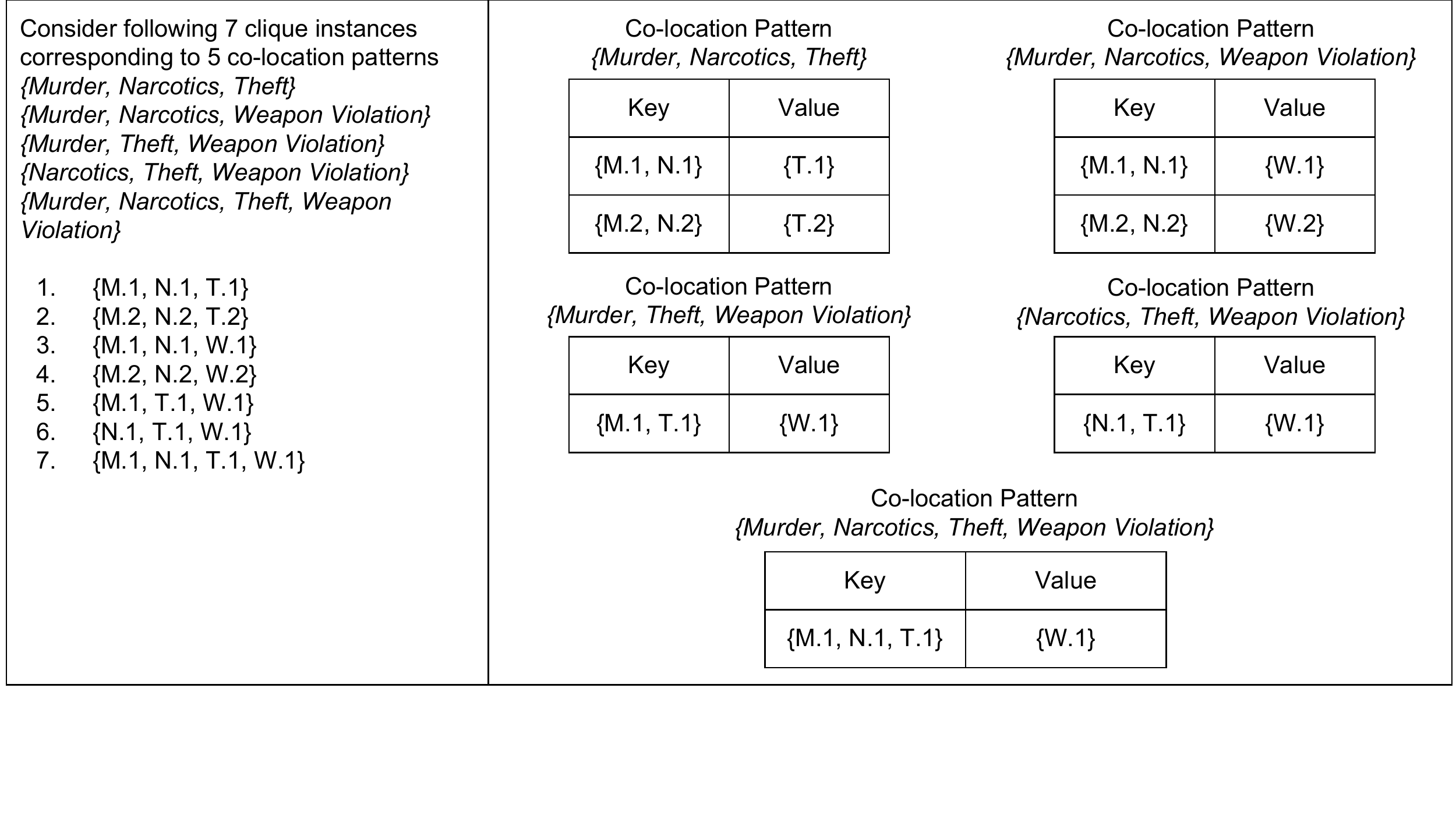}
\end{table}

\subsection{CliqueExtend Algorithm}
\label{subsection:cliqueExtend}
\begin{algorithm}
\begin{algorithmic}[1]
\caption{CliqueExtend Algorithm}
\label{alg4}
\REQUIRE Neighborhood Graph $G(V, E)$, $minPrev$, $k$
\ENSURE $prevalentColocations$ - a key-value store where the value for key $k'$ corresponds to an ordered set of prevalent co-locations of size $k'$
\STATE $prevlanetColocations[1] = new OrderedSet()$
\FOR{vertex $v \in V$} 
\STATE $prevalentColocations[1].insert(v.feature)$
\ENDFOR
\STATE $Mcurrent$ = new $KeyValueStore()$ 
\STATE $Mprevious$ = new $KeyValueStore()$ 
\FOR {$k`$ in $(2, 3, \cdots, k)$} 
\STATE $candidateColocations =$
\item[] $aprioriGen(prevalentColocations[k'-1])$
\STATE $prevalentColocations[k']$ = new $OrderedSet()$
\STATE $Mcurrent.clear()$
\FOR{$candidate \in candidateColocations$}
\STATE $candidateCliques =$
\item[] $generateCliques(Mprevious, candidate)$
\STATE $Mtemp$ = new $KeyValueStore()$ 
\FOR{$candidateClique \in candidateCliques$}
\IF{$lookupEdge(G, candidateClique)$}
    \STATE $Mtemp.insert(candidateClique)$
\ENDIF
\ENDFOR
\IF{$prevalance(candidate) \geq minPrev$}
    \STATE $prevalentColocations[k'].append(candidate)$
    \STATE $add(Mtemp, Mcurrent$)
\ENDIF
\ENDFOR 
\STATE $Mprevious = Mcurrent$
\ENDFOR
\end{algorithmic}
\end{algorithm}
CliqueExtend is a partial traversal based algorithm as only the second step, i.e., validation of candidate clique instances involves traversal on $G$. First step of candidate clique enumeration of size $k$ is performed by extending clique instances of size $k-1$ stored in key-value store. \cut{In this algorithm, size $k-1$ clique instances are stored and used to enumerate size $k$ candidate clique instances. }Explanation of the detailed steps of the Algorithm \ref{alg4}:

\textbf{Line 1-11} remains same as for Algorithm \ref{alg3}. \textbf{Line 12} \cut{Candidate clique instances are enumerated using the prevalent clique instances stored in $Mprevious$. }We propose a new technique for generating size $k$ candidate clique instance from two $k-1$ clique instances stored in $Mprevious$. Consider Table \ref{table:DataStructure} for following example. We want to enumerate candidate clique instances for SCP $X = \{Murder, Narcotics, Theft, Weapon\ Violation\}$. We consider clique instances of SCP $X1 = \{Murder, Narcotics, Theft\}$ and $X2 = \{Murder, Narcotics, Weapon\ Violation\}$. We have key $\{M.1, N.1\}$ present in key-value stores of both SCPs $X1$ and $X2$, thus, candidate clique instance(s) for SCP $X$ is(are) $\{M.1, N.1, T.1, W.1\}$. So this way we can enumerate all possible candidate clique instances for size $k$ using two clique instances of size $k-1$. This logic is encoded in the $generateCycles$ method.

\textbf{Line 13-25} remains same as for Algorithm \ref{alg2} with one modification. Instead of using the $validateClique$ method (from Algorithm \ref{alg2}), the $lookupEdge$ method is used to validate if the $candidateClique$ is indeed a clique. \cut{This method performs an edge look-up in the neighborhood graph to see if the two vertex ids corresponding to a common key in $Mprevious$ are connected. Continuing with the example we saw in last step, we would check if $T.1$ and $W.1$ are connected.} Just like the previous two algorithms, clique enumeration (line 12) and clique validation (line 5) can be executed in parallel and Algorithm \ref{alg3} can also be scaled horizontally.

\section{Experimental Setup and Results}
\label{sec:Results}

We used real world crime dataset of City of Chicago, USA \cite{dataset} for all our experiments. The data consists of crime incidents with primary crime type, address of crime incident (lat and long), date and time when crime incident occurred. We used data corresponding to 30 thousand incidents spread across 33 distinct crime types. 
\cut{\begin{figure}
\center
\includegraphics[width=0.75\linewidth]{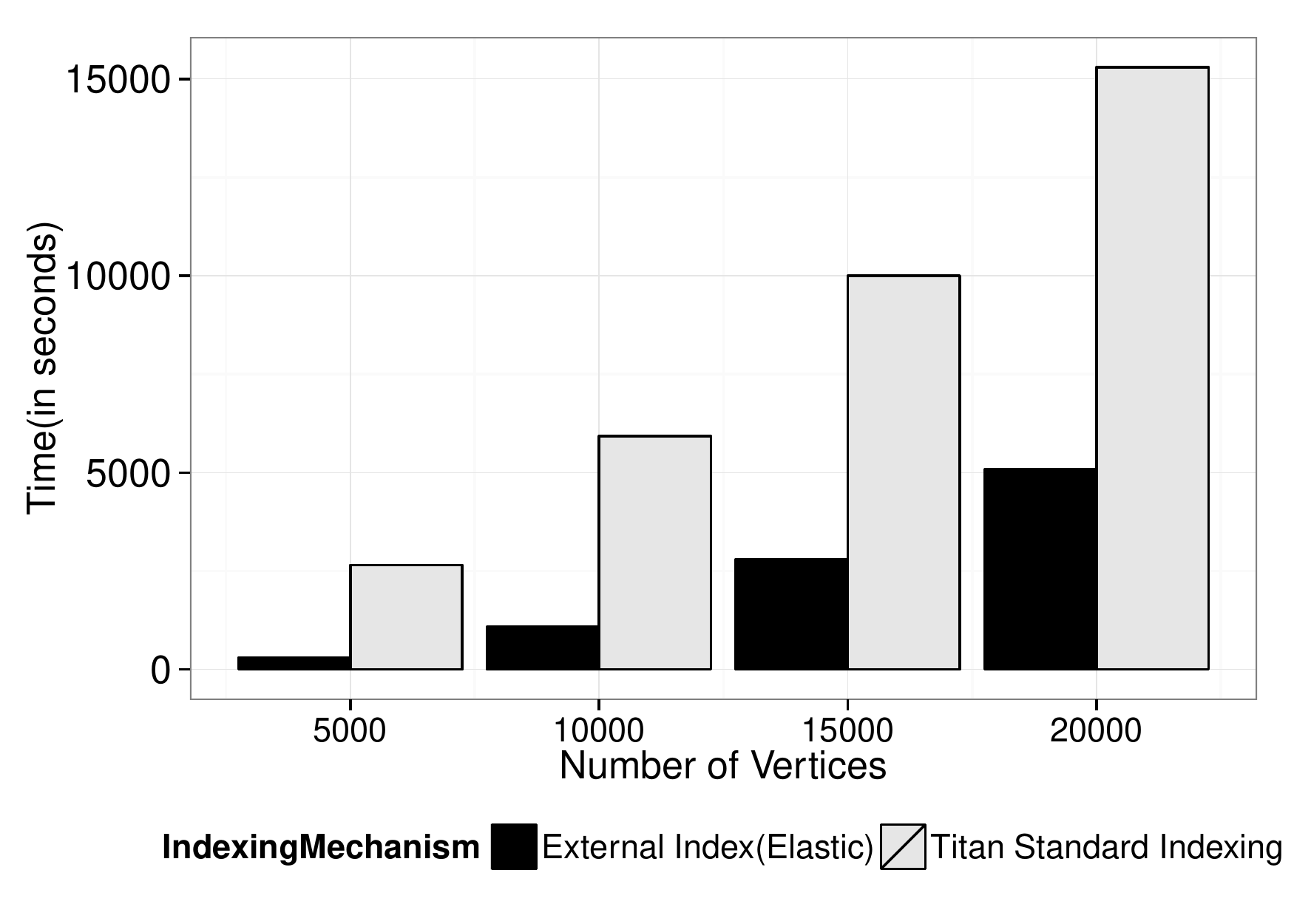}
\caption{Neighborhood exploration time using external indexing (Elastic) and Titan standard indexing}
\label{fig:indexing}
\end{figure}}
We used Titan \cite{titan}, a scalable, distributed graph database, to materialize and store the neighborhood graph. We use Titan with the Cassandra as our back-end store as most of our queries are read queries. It ensures availability, partition tolerance and eventual consistency. We partition the graph using the edge cut strategy to minimize internode communication during edge traversal. We use MongoDB as our key-value store as it supports multi-granularity locks at global, database and collection level. This level of granularity is crucial for our algorithms to execute in parallel so that the read and write operations for different candidate co-locations do not block on each other. We use Elastic \cut{\footnote{https://www.elastic.co/}} cluster to index Titan. 

\begin{figure}
\center
\includegraphics[width=0.75\linewidth]
{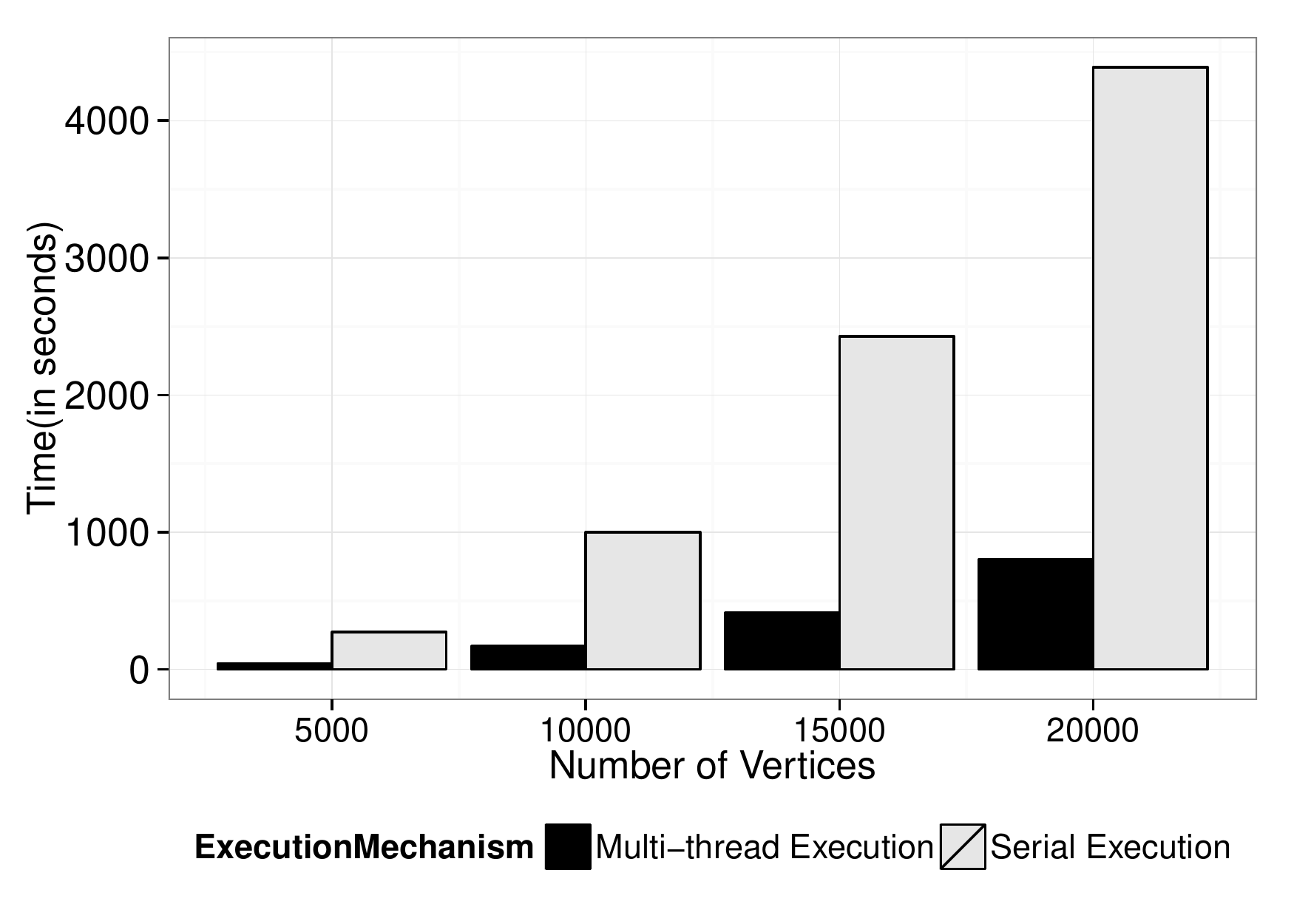}
\caption{Edge insertion time in titan graph using serial execution and multi-threaded execution}
\label{fig:result3}
\end{figure}

In Figure \ref{fig:result3} we compare the edge insertion time using single-threaded and multi-threaded implementation. Since Elastic is distributed in nature, multi-threaded implementation beats single-threaded implementation.
Finally, we compare neighborhood exploration time in following scenarios : 
\begin{enumerate}
\item Neighborhood Exploration using Edge Traversal
\item Neighborhood Exploration using Single-threaded exectution of geo-range query
\item Neighborhood Exploration using Multi-threaded exectution of geo-range query
\end{enumerate}
We observe (Table \ref{table:neighborhood_exploration}) that edge traversal is orders of magnitude faster than both kind of geo-range query. This provides the motivation for inserting edges in the neighborhood graph instead of using geo-range queries every time.

\begin{table}
\centering
\caption{Number of Vertices vs Neighborhood Exploration Time (in seconds) 
\label{table:neighborhood_exploration}}
\begin{tabular}{|c|c|c|c|}
\hline 
\thead{N} & \thead{Edge Traversal\\ (Single-Threaded)}  & \thead{Elastic\\(Multi-Threaded)} & \thead{Elastic\\(Single-Threaded)}\\
\hline
5000 & 0.136 & 5.307 & 298.019 \\
\hline
10000 & 0.269 & 7.484 & 1088.636 \\
\hline
15000 & 0.375 & 11.926 & 2797.643 \\
\hline
20000 & 0.494 & 16.526 & 5093.748 \\
\hline
\end{tabular}
\end{table}

Now we report the run time analysis results of our proposed algorithms. We have three user-defined parameters as shown in Table \ref{table:summary_params}. For experiments in this section, we fix two of the three parameters to their default value and vary the remaining parameter over its range. For all these experiments, we have implemented multi-threaded version of proposed algorithms with the Java thread pool size set to 48.
\begin{table}
\centering
\caption{Summary of parameters
\label{table:summary_params}}
\begin{tabular}{|c|c|c|c|}
\hline 
Parameters & Notation & Range & Default Value\\
\hline
Number of Vertices & N & $[10^4, 3*10^4]$ & $10^4$ \\
\hline
Threshold Distance & R & [0.3, 0.5]km & 0.3km \\
\hline
\makecell{Threshold Participation \\Index (min\_prev)} & Threshold PI & [0.01, 0.1] & 0.1 \\
\hline
\end{tabular}
\end{table}
\begin{table}
\centering
\caption{Variation of time taken (in seconds) by all the three algorithms to generate SCPs till size $4$ vs. N.\cut{ R and PI set to default}
\label{table:result_time_vs_vertices}}
\begin{tabular}{|c|c|c|c|}
\hline 
N & CliqueExtend & CliqueEnum\textsubscript{G} & CliqueEnum\textsubscript{K} \\
\hline
$1*10^4$ & 27.145 & 34.374 & 53.081 \\
\hline
$2*10^4$ & 116.799 & 191.315 & 1266.695 \\
\hline
$3*10^4$ & 392.437 & 578.46 & 12837.177 \\
\hline
\end{tabular}
\end{table}
\subsection{Varying the number of vertices (N) in $G$}
Table \ref{table:result_time_vs_vertices} shows the variation of time taken (in seconds) to generate co-location patterns till size 4 vs. N . We increase N in steps of $10^4$. As N increases, the time taken by all the three algorithms increases. The rationale behind this observation is that with increasing number of vertices, we have more neighbors to enumerate and more candidate clique instances to validate as compared to case where the number of vertices is less. Observe that CliqueExtend algorithm performs consistently better than the CliqueEnum\textsubscript{G} which in turn performs better than CliqueEnum\textsubscript{K}. Also note that as N increases, the performance difference between the three approaches increases, making CliqueExtend the clear winner.
\begin{table}[h]
\centering
\caption{Variation of time taken (in seconds) by all the three algorithms to generate SCPs till size $4$ vs. PI.
\label{table:result_time_vs_pi}}
\begin{tabular}{|c|c|c|c|}
\hline
PI & CliqueExtend & CliqueEnum\textsubscript{G} & CliqueEnum\textsubscript{K} \\
\hline
0.01 & 59.135 & 88.210 &  103.510 \\
\hline
0.05 & 38.629 & 50.293 & 88.952 \\
\hline
0.1 & 27.145 & 34.374 & 53.081 \\
\hline
\end{tabular}
\end{table}
\subsection{Varying the threshold participation index (Threshold PI)}
Table \ref{table:result_time_vs_pi} shows the variation of time taken (in seconds) to generate SCPs till size $4$ vs. threshold PI. As threshold PI increases, time taken by all the three algorithms decreases. With increasing value of threshold PI, lesser number of SCPs will be prevalent. So at each iteration we have fewer candidate SCPs as compared to the case where value of threshold PI is lower. Observe that CliqueExtend algorithm performs better than CliqueEnum\textsubscript{G} which in turn beats CliqueEnum\textsubscript{K}.
\begin{table}
\centering
\caption{Variation of time taken (in seconds) by all the three algorithms to generate SCPs till size 4 vs. R.
\label{table:result_time_vs_R}}
\begin{tabular}{|c|c|c|c|}
\hline
R (in km) & CliqueExtend & CliqueEnum\textsubscript{G} & CliqueEnum\textsubscript{K} \\
\hline
0.3 & 27.145 & 34.374 & 53.081 \\
\hline
0.4 & 52.070 & 65.001 & 367.736 \\
\hline
0.5 & 83.25 & 101.798 & 1247.796 \\
\hline
\end{tabular}
\end{table}
\subsection{Varying the threshold distance (R)}
Table \ref{table:result_time_vs_R} shows the variation of time taken to generate SCPs till size $4$ vs. R. As R increases, the time taken by all the three algorithms increases as the number of edges in $G$ increases. CliqueExtend algorithm performs better than CliqueEnum\textsubscript{G} which is better than the CliqueEnum\textsubscript{K}.

\begin{table}
\centering
\caption{Variation of time taken (in seconds) by all the three algorithms to SCPs vs. size of SCP. Threshold PI = 0.01
\label{table:result_time_vs_size}}
\begin{tabular}{|c|c|c|c|}
\hline
\thead{Size of SCP} & CliqueExtend & CliqueEnum\textsubscript{G} & CliqueEnum\textsubscript{K} \\
\hline
2 & 6.079 & 5.680 & 6.033 \\
\hline
3 & 47.201 & 42.037 & 46.634 \\
\hline
4 & 59.135 & 89.21 & 103.51 \\
\hline
5 & 70.123 & 128.455 & 184.914 \\
\hline
6 & 78.062 & 152.798 & 273.951 \\
\hline
7 & 80.8 & 162.975 & 343.379 \\
\hline
\end{tabular}
\end{table}
\subsection{Run time analysis corresponding to each iteration}
Table \ref{table:result_time_vs_size} shows the variation of time taken (in seconds) to generate SCP of varying sizes for the three different algorithms. As the SCP size increases, time taken by CliqueEnum\textsubscript{G} and CliqueEnum\textsubscript{K} increases significantly as compared to CliqueExtend algorithm. For size $2$ and $3$, time taken by CliqueExtend is slightly more than other algorithms which can be explained by write overhead associated with services like MongoDB. CliqueExtend algorithm performs better than CliqueEnum\textsubscript{G} which is better than CliqueEnum\textsubscript{K}.
\subsection{Dynamic Neighborhood Constraint}
Table \ref{table:result_create_vs_update1} shows the variation of time taken for constructing a new graph vs. the time taken to construct incrementally upon the existing graph (update) when the threshold distance changes. N is fixed at $20,000$. The threshold distance is updated in constant steps of size $0.1$ km. The updation time for R = $0.5$ km is the time taken to construct the graph with threshold distance of $0.5$ km from existing graph with its distance threshold set as $0.4$ km. Note that graph updation time is much less than graph creation time and we can vary the threshold distance to 0.5 km without doing all the computations again. This enables to perform interactive analysis by varying threshold distance.
\begin{table}
\centering
\caption{Comparison of time taken for graph construction vs. incremental graph construction when threshold distance changes in steps of 0.1 km
\label{table:result_create_vs_update1}}
\begin{tabular}{|c|c|c|}
\hline
R (in km) & Creation Time (in seconds) & Updation Time (in seconds)\\
\hline
0.5 & 313.825 & 111.412  \\
\hline
0.6 & 439.217 & 117.653 \\
\hline
0.7 & 621.309 & 128.194 \\
\hline
0.8 & 786.542 & 149.729 \\
\hline
0.9 & 983.707 & 160.57 \\
\hline
\end{tabular}
\end{table}

\textbf{Discussion}
The general observation in terms of performance is CliqueExtend $>$ CliqueEnum\textsubscript{G} $>$ CliqueEnum\textsubscript{K}.

For discovery of size $k$ SCPs, both CliqueEnum\textsubscript{G} and CliqueEnum\textsubscript{K} traverse graph to generate the candidate clique instances (cycles in these two cases). For validating these instances for cliques, CliqueEnum\textsubscript{G} performs $O(k^2)$ edge look-ups on $G$ while CliqueEnum\textsubscript{K} performs only 2 look-up over a MongoDB database. The multiple edge look-ups outperform the 2 look-ups over MongoDB primarily because in the case of MongoDB, the key is first $k-1$ vertices of the clique instance while in the case of $G$, the look-ups are only size $1$ elements of the clique instance. In case of $CliqueExtend$ and $CliqueEnum_G$, $CliqueExtend$ benefits from faster clique validation. For $CliqueEnum_G$, clique candidate instances are cycles and hence $O(k^2)$ edge look-ups are needed. But in the case of CliqueExtend, candidate cliques are more strongly connected than the case of cycles and a single edge look-up is sufficient to validate whether the candidate is a clique or not. Notice that while CliqueExtend is much faster than CliqueEnum\textsubscript{G}, it also needs more storage as it needs to store all the size $k-1$ clique instances (which are used for generating size $k$ clique instances) unlike CliqueEnum\textsubscript{G} which generates clique instances using graph traversal. In that way CliqueExtend provides a memory-speed tradeoff.

Our algorithms support interactive user analysis based on varying distance threshold as graph update works orders of magnitude faster than graph creation.


\section{Conclusion}
\label{sec:Conclusion}
We present a novel perspective to SCP mining - ``Developing techniques for SCP mining using Graph Database". We introduced the concept of neighborhood graph and modeled it as a property graph which we materialize using Titan graph database. We proposed three algorithms for SCP mining using graph database - CliqueEnum\textsubscript{G}, CliqueEnum\textsubscript{K} and  CliqueExtend.  We implemented a multi-threaded version of proposed algorithms and our 
results established that CliqueExtend performs the best
followed by CliqueEnum\textsubscript{G} and CliqueEnum\textsubscript{K}.

Our algorithm supports interactive-user analysis and the 
neighborhood constraint parameters can be varied over a range.

We leveraged a key-value store to either enumerate candidate cliques or to validate them - but not for both. Exploring the possibility of a fully key-value store based approach is part of future work. Here, we focused on the spatial aspect of SCP mining. A natural extension  would be in the domain of spatial data mining wherein our neighborhood graph can be leveraged to suit the requirements of respective domains. \cut{We also plan to augment dynamic distance threshold for enhanced interactive-user analysis.}



\bibliographystyle{abbrv}
\bibliography{ref}  

\end{document}